\documentclass[a4paper]{article} 

\usepackage{amsmath}
\usepackage{amsfonts}
\usepackage{book_pc_no_env}
\usepackage{code}
\usepackage{fancyvrb}
\usepackage{pifont}
\usepackage{draftwatermark}
\SetWatermarkScale{5}

\newtheorem{theorem}{Theorem}
\newtheorem{corollary}[theorem]{Corollary}

\newenvironment{proof}
	{\par\noindent\upshape\textbf{Proof}\quad}
	{\hspace*{\fill}$\Box$}

\title{Confusion in the Church-Turing Thesis %\\ DRAFT: NOT FOR CIRCULATION
}
\author{Barry Jay and Jose Vergara \\
 University of Technology, Sydney \\
\{Barry.Jay,Jose.Vergara\}@uts.edu.au}

\begin{document}

\maketitle
\makeatactive

\begin{abstract}

  The Church-Turing Thesis confuses numerical computations with
  symbolic computations. In particular, any model of computability in
  which equality is not definable, such as the $\lambda$-models
  underpinning higher-order programming languages, is {\em not}
  equivalent to the Turing model. However, a modern combinatory
  calculus, the $SF$-calculus, can define equality of its closed
  normal forms, and so yields a model of computability that is
  equivalent to the Turing model. This has profound implications for
  programming language design.
\end{abstract}

\section{Introduction}
\label{sec:intro}

The $\lambda$-calculus \cite{Church41,Bare84a} does not define the limit of
expressive power for higher-order programming languages, even when
they are implemented as Turing machines \cite{Turing36}. That this
seems to be incompatible with the Church-Turing Thesis is due to
confusion over what the thesis is, and confusion within the thesis
itself.  According to Soare \cite[page 11]{soare1999history}, the
Church-Turing Thesis is first mentioned in Steven Kleene's book {\em
  Introduction to Metamathematics} \cite{Kleene52}. However, the
thesis is never actually stated there, so that each later writer has
had to find their own definition. The main confusions within the
thesis can be exposed by using the book to supply the premises for the
argument in Figure~\ref{fig:faulty} overleaf. The conclusion asserts
the $\lambda$-definability of equality of closed $\lambda$-terms in normal form,
i.e.\ syntactic equality of their deBruijn representations
\cite{deBruijn}.  Since the conclusion is false \cite[page
519]{Bare84a} there must be a fault in the argument. 

The basic error is introduced by statement $(2)$, since Turing's proof
was for {\em numerical} functions, i.e.\ functions acting on natural
numbers or positive integers, and not for {\em symbolic} functions,
i.e.\ functions acting on words in some alphabet, usually well-formed
formulas, such as $\lambda$-terms.  So, there is a confusion between
numerical interpretation and symbolic interpretation. Since the link
to higher-order programming concerns symbolic functions, we should
replace $(2)$ by a less ambiguous statement, such as:

\begin{description}
\item[$(2')$]   The computable symbolic functions are equivalent to
the $\lambda$-definable symbolic functions.
\end{description}

\noindent 
Now the focus shifts to the nature of the equivalence in $(2')$ and
its use when inferring $(3)$.  We will see that the traditional
arguments do indeed induce a relationship between a $\lambda$-model of
computability and a Turing model, in that each can simulate the other,
but this mutual simulation is too weak to be an equivalence, to support
the substitution of ``$\lambda$-definable'' for ``computable'' in the argument.

\begin{figure}
\begin{tabular}{rlr}
$(1)$ &
 ``... every function which would naturally be
  regard- \\ & ed as computable is computable under his [Turing's] \\ &  definition, i.e. by
  one of his machines ...'' & (page 376) \\
$(2)$ &  ``The equivalence of the computable to the $\lambda$-\\ &  definable functions \ldots was proved by Turing 1937.'' & (page 320) \\
$(3)$ &
  Every function which would naturally be regarded  \\ & as computable is $\lambda$-definable. &(equivalence) \\
$(4)$ &
The equality of  $\lambda$-terms in closed normal form is \\ & $\lambda$-definable. &(specialise) 
\end{tabular}
\caption{A faulty argument, with premises quoted from Kleene \cite{Kleene52}}
\label{fig:faulty}
\end{figure}

Thus we have three confusions. The faulty argument exposes the
confusion of numerical functions with symbolic ones, and of mutual
simulation with equivalence. Then there is the confusion about what,
exactly, the Church-Turing Thesis is.

We will explore these confusions in three stages.  First, we will
explore the confusion surrounding Church's Thesis, as introduced by
Kleene \cite{Kleene52}. This will lead us to formalise models of
computability and their relative expressive power. It will follow that
any $\lambda$-model of computability that is suitable for modeling
higher-order programming languages is less expressive than the
recursive model.  Second, we will extend the analysis to Turing's
Thesis and the Church-Turing Thesis, and their impact on programming
language design.  Third, we will show that a recent combinatory
calculus, the $SF$-calculus \cite{JGW11} yields a model of
computability that is both suitable for higher-order programming and
equivalent to the Turing model.  By supporting a form of intensional
computation, it suggests a more powerful approach to programming
language design.

%\subsection{Structure of the paper}  

The sections of the paper are: Section~\ref{sec:intro} Introduction;
Section~\ref{sec:Church} Church's Thesis; Section~\ref{sec:models}
Models of computability; Section~\ref{sec:equiv_models} Comparison of
models; Section~\ref{sec:Turing} Turing's Thesis;
Section~\ref{sec:Church-Turing} The Church-Turing Thesis;
Section~\ref{sec:design} Programming language design;
Section~\ref{sec:intensional} Intensional computation; and
Section~\ref{sec:conclusions} Conclusions.

\section{Church's Thesis} 
\label{sec:Church}

The background to Church's Thesis is Church's paper of 1936 {\em An
  unsolvable problem of number theory} \cite{Church36} which,
alongside Alan Turing's work discussed later, showed that some
numerical problems do not have a computable solution, so that Hilbert's
decision problem does not have a general solution.  It is clear that
Church's focus was on numerical functions, as all his formal
definitions were expressed in numerical terms. For example, he writes,
``A function F of one positive integer is said to be $\lambda$-definable if
\ldots'' \cite[page 349]{Church36}. Again, in Theorem~XVII he writes
``Every $\lambda$-definable function of positive integers is recursive''.

That said, he does broaden the discussion
when considering {\em effective calculability}. On the one hand, an
{\em effectively calculable} function of positive integers is defined
to be ``a recursive functions of positive integers'' or ``a
$\lambda$-definable function of positive integers''. On the other hand, he writes:
\begin{quote}
\ldots [in] some particular system of symbolic logic \ldots it is
necessary that each rule of procedure be an effectively calculable
operation, \ldots\ Suppose we
interpret this to mean that, in terms of a system of G\"odel
representations for the expressions of the logic, each rule of
procedure must be a recursive operation, \ldots
\end{quote}
That is to say, a symbolic function is recursive if it can be
simulated by a numerical function that is recursive, where simulation
is defined using G\"odel numbering.  Further, the difference between the
numerical and symbolic domains does not seem to be important to him,
as he writes, in a footnote:
\begin{quote}
  \ldots in view of the G\"odel representation and the ideas
  associated with it, symbolic logic can now be regarded,
  mathematically, as a branch of elementary number theory.
\end{quote}

In contrast to Church's narrow constructions, Kleene's definitions
have a broad scope. For example, he gives three definitions of
$\lambda$-definability in his paper {\em $\lambda$-definability and
  recursiveness} \cite{Kleene36}, according to whether the domain of
definition is the non-negative integers, the $\lambda$-terms themselves, or
other mathematical entities for which a $\lambda$-representation has
been given. It is the interplay between these definitions that is the
primary cause of confusion.

Kleene introduces Church's Thesis as Thesis I in his 1952 book 
(Section~60, page 300) as follows:

\begin{description}
\item[(CT)] Every effectively calculable function (effectively decidable predicate) is general recursive. 
\end{description}

\noindent
The crucial question is to determine the domain of the effectively
calculable functions.  

At the point where Church's Thesis is stated, in Section~60, the
general recursive functions are all numerical, so it would seem that
the effectively calculable functions must also be numerical.  However,
he does not include the phrase ``of positive integers'' in his
statement of the thesis, in the careful manner of Church. We are
required to add this rider in order to make sense of the thesis.

Later, in Section~62.\ Church's Thesis, Kleene presents seven pages of
arguments for the thesis, which he groups under four headings A--D.  In
``(B) Equivalence of diverse formulations'' he asserts that the set of
$\lambda$-definable functions and the set of Turing computable functions
are ``co-extensive'' with the set of general recursive functions.
Again, this only makes sense if the functions are presumed to be
numerical functions. This paragraph is also the source of statement
$(2)$ from Figure~\ref{fig:faulty}.

If we were to stop at this point, then the explanation of the faulty
argument would be quite simple: statement $(2)$ should be read in a
context where all functions are numerical, or be replaced by ``The
equivalence of the computable to the $\lambda$-definable {\bf numerical}
functions was proved by Turing 1937''. The restriction to numerical
functions propagates to statement (3) which cannot then be specialised
to equality of $\lambda$-terms. 

However, in ``(D) Symbolic logics and symbolic algorithms'', Kleene
reprises Church's definition of symbolic functions that are recursive,
so that, by the end of Section~62, we have two definitions of
effectively calculable functions and of recursive functions. So there
are two versions of Church's Thesis, one for numerical functions (NCT)
and one for symbolic functions (SCT).  Unpacking the definition of a
general recursive symbolic function to make the role of simulations
explicit, the two versions become:

\begin{description}
\item [(NCT)]  Every effectively calculable numerical function (effectively decidable numerical predicate) is general recursive. 
\end{description}

\begin{description}
\item [(SCT)]  Every effectively calculable function (effectively decidable predicate) can be simulated by a function which is general recursive. 
\end{description}

Summarising, we see that Church was careful to separate theorems about
numerical functions from the discussion of computation in symbolic
logic. By contrast, Kleene presents a single statement of Church's
Thesis with evidence that confuses the numerical with the symbolic.
In turn, this confuses two different questions: whether two sets of
numerical functions are the same; and whether there is an encoding
that allows functions in a symbolic logic to be simulated by recursive
functions on numbers.  These confusions can be defused by isolating
two versions of Church's Thesis which, from the viewpoint of Post
\cite{post}, qualify as scientific laws. Now this distinction is
enough to eliminate the numerical version of the faulty argument but
the symbolic version remains.  To eliminate it, we must show that an
equivalence of models of computation requires more than their mutual
simulation.  To make this precise, we must formally define models of
computability, their simulations and equivalence.

\section{Models of computability} 
\label{sec:models}

We adopt the simplest definition of model of computability in which
the discussion of simulation makes sense. This was introduced by
Boker and Dershowitz as a {\em model of computation} \cite{BokerD06},
but since the focus is upon functions for which a computation is
possible, rather than the actual mechanics of computation, it seems
more accurate to call them models of computability. Note, too, that
the domain of the computable functions is not actually required to be
symbolic in any way, or even to be enumerable, however natural this
may be. This makes the notion too weak for many purposes, but here it
emphasises the generality of our results. 

A {\em model of computability} $(D, {\cal F})$ is given by a domain
$D$ which is a set of {\em values} that provides arguments and results
of computations, and a collection ${\cal F}$ of partial functions from
powers of $D$ to $D$.  Here are some examples.

The partial recursive functions on natural numbers form a model with
domain given by the natural numbers and functions given by
the partial recursive functions. Call this the {\em recursive model of
  computability}.

Recall that an injective function is a total function that does not
identify distinct arguments.  For any finite alphabet $\Sigma$, and any
domain $D$ equipped with an injective function from $D$ to the words
of $\Sigma$, the {\em Turing model of computability on $D$} has domain
$D$ and partial functions given by those which can be computed by a
Turing machine with alphabet $\Sigma$.  When the choice of domain is
understood from the context, or unimportant, then it may be called the
{\em Turing model of computability}.

Any {\em applicative rewriting system} \cite{TRS} 
has a {\em normal model} whose values are the closed terms in normal form,
and whose partial functions are those representable by  closed terms
in normal form. Further, any subset $D$ of values determines a model,
where the partial functions are now defined on a value in $D$ only if
the result is also in $D$. 

For example, classical combinatory logic has terms built from
applications of the operators $S$ and $K$ and variables.  As well
as its normal model, there is a {\em numerical model} whose domain is
restricted to be the Church numerals. Also, one can use Polish
notation to encode combinators as words using the alphabet $\Sigma =
\{A,S,K\}$, where $A$ is for application. For example, $S(KK)$ is
mapped to $ASAKK$. This yields a Turing model of computability for
$SK$-normal forms.

Similarly, $\lambda$-calculus has {\em normal models} and {\em numerical
  models}, once the terms and reduction rules have been specified.
First, a $\lambda$-term is unchanged by renaming of bound variables, i.e.\
is an $\alpha$-equivalence class in the syntax \cite{Bare84a}. Since
this equivalence is a distraction, we will work with $\lambda$-terms using
deBruijn notation \cite{deBruijn} so that, for example, $\lambda x.x $
becomes $\lambda 0$ and $\lambda x. \lambda y. x y$ becomes $\lambda \lambda 1 0$. Second,
there are various choices of reduction rules possible, with each
choice producing a normal $\lambda$-model.  Define a {\em $\lambda$-model of
  higher-order programming} to be any normal $\lambda$-model in which
equality of closed normal forms is not definable.  This excludes
any model whose domain is numerical, and would seem to include any models
that are relevant to higher-order programming.  Certainly the
$\lambda$-models of higher-order programming include those given by
$\beta$-reduction alone, or $\beta\eta$-reduction.

\section{Comparison of models} 
\label{sec:equiv_models}

Now consider what it means for one model of computability to be more
expressive than another.  The simple interpretation requires that
their computable functions have the same domain and then compares sets
of computable functions by subset inclusion, as is done by John
Mitchell \cite{Mitchell92} and Neil Jones \cite{Jones97}.  The choice
of domain is important here. For example, if the domain consists of
natural numbers then the $\lambda$-model and the recursive model are indeed
equivalent.  However, this restriction is unreasonable for modeling
higher-order programming since functions must be among the values. Now
it is an easy matter to see that any normal $\lambda$-model of
computability has fewer computable functions than the Turing model.
For example, no $\lambda$-term can decide equality of values but this
function is in the Turing model.

The complex interpretation of relative expressive power allows the
domains to vary, but now the comparison of computability over domains
$D_1$ and $D_2$ must be mediated by simulations, which are given by
encodings.  Note that it makes no sense to consider a simulation in
one direction, and compare sets of functions in the other direction,
since, as observed by Boker and Dershowitz \cite{BokerD06}, this can
lead to paradoxes.  Rather, there should be encodings of each domain
in the other that are, in some sense, inverse to each other. Various
choices are possible here but two requirements seem to be essential.
First, the encodings should be injective functions, since distinct
values should not be identified.  Second, the encodings should be {\em
  passive}, in the sense that they are not adding expressive power
beyond that of their target model. In particular, they should be
effectively calculable.  This requirement can only be verified
informally, on a case by case basis, since functions from $D_1$ to
$D_2$ are not in the scope of either model.

A {\em simulation} of a model of computability $(D_1,{\cal F}_1)$ in another such
$(D_2, {\cal F}_2)$ is given by an injective {\em encoding} $\rho : D_1
\to D_2$ such that every function $f_1$ in ${\cal F}_1$ can be {\em
  simulated by} a function $f_2$ in ${\cal F}_2$ in the following sense:
for all $x_1,\ldots, x_n \in D_1$ such that $f_1(x_1,\ldots, x_n)$ is
defined, then $f_2(\rho(x_1),\ldots, \rho(x_n))$ is defined and
\[
\rho(f_1(x_1,\ldots, x_n)) = f_2(\rho(x_1),\ldots, \rho(x_n)) \; . 
\]

For example, G\"odelisation provides a simulation of the normal
$SK$-model into the recursive model. More generally, Church and Kleene
both use this notion of simulation to define  recursive
symbolic functions. 
G\"odelisa\-tion seems to be passive.

Further, the encoding of the natural numbers using Church numerals
provides a simulation of the recursive model into the normal
$SK$-model or any normal $\lambda$-model.

Other related notions of simulation can be found in the literature,
e.g.\ \cite{Jones97, BokerD06}.  For example, Richard Montague
\cite[page 430]{Montague60} considers, and Hartley Rogers \cite[page
28]{Rogers67} adopts, a slightly different approach, in which the
encoding of numbers is achieved by reversing a bijective
G\"odelisation.  However, this inverse encoding may not be a
simulation.  For example, the equality of numbers cannot be simulated
by a $\lambda$-term over the domain of closed $\lambda$-terms in normal form.
Rogers, like Kleene, ensures a simulation by {\em defining} the
computable functions in the symbolic domain to be all simulations of
partial recursive functions, but this says nothing about
$\lambda$-definability.

Given the formal definition of simulations, it may appear that the
strongest possible notion of equivalence is that each model simulates
the other. However, if the encodings are passive then so are the {\em
  recodings} from $D_1$ to $D_1$ and from $D_2$ to $D_2$ obtained by
encoding twice.  Since these are in the scope of the two models, we
can require that recodings be computable.

Let $(D_1, {\cal F}_1)$ and $(D_2, {\cal F}_2)$ be two models of
computability with simulations $\rho_2 : D_1 \to D_2$ and $\rho_1 :
D_2 \to D_1$. Then $(D_2,{\cal F}_2)$ is {\em at least as expressive} as
$(D_1,{\cal F}_1)$ if the recoding $\rho_2\circ\rho_1 : D_2 \to D_2$
is computable in ${\cal F}_2$. If, in addition, $(D_1,{\cal F}_1)$ is
more expressive than $(D_2,{\cal F}_2)$ then the two models are {\em
  weakly equivalent}. 
Note that this notion of equivalence is indeed an equivalence
relation on models of computability.

It is interesting to compare this definition with those for
equivalence of {\em partial combinatory algebras} by Cockett and
Hofstra \cite{CH2010} and John Longley \cite{Longley14}.  They would
not require the encodings to be injective, but Longley
would require that the recodings be invertible. Adding the latter
requirement is perfectly reasonable but is immaterial in the current
setting.

It is easy to prove that the recursive model is at least as expressive as
any $\lambda$-model. Our focus will be on the converse. 

\begin{theorem}
  Any model of computability that is at least as  expressive as the
  recursive function model can define equality of values.
\end{theorem}

\begin{proof}
  The recursive model is presumed to use $0$ and $1$ for booleans.  In
  the other model, identify the booleans with the encodings of $0$ and
  $1$ so that the equality function is given by recoding its arguments
  and then applying the simulation of the equality of numbers.
\end{proof}

\begin{corollary}
\label{cor:SK-inequivalent}
The normal model of computability for $SK\!$-calculus is not weakly equivalent to the recursive
function model.
\end{corollary}

\begin{proof}
  If the normal $SK$-model could define equality then it could distinguish
  the values $SKK$ and $SKS$ but the standard translation from
  combinatory logic to $\lambda$-calculus identifies them (both reduce to
  the identity), and so they cannot be distinguished by any
  $SK$-combinator. 
\end{proof}

\begin{corollary}
\label{cor:lambda-inequivalent}
No $\lambda$-model of higher-order programming is weakly equivalent to the
recursive model of computability.
\end{corollary}

\begin{proof}
  Since normal $\lambda$-models do not define equality, the result is
  immediate. Note that Longley has proved the analogous result for his
  definition of equivalence \cite{Longley14}.
\end{proof}

\section{Turing's Thesis}  
\label{sec:Turing}

Turing's paper of 1936 {\em On Computable Numbers, with an application
  to the Entscheidungsproblem} was, like Church's paper, concerned
with numerical computation and Hilbert's decision problem. Like
Church, Turing was careful to limit his definitions, e.g.\ of {\em
  computable functions}, to numerical functions while showing
awareness of a broader scope. For example, in the first paragraph he
writes:
\begin{quote}
  Although the subject of this paper is ostensibly the computable {\em
    numbers}, it is almost equally easy to define and investigate
  computable functions of an integral variable, or a real or
  computable variable, computable predicates, and so
    forth.
\end{quote}
Similarly, the use of an unspecified
  alphabet of symbols on the tape of a Turing machine encourages us to
  consider computation over arbitrary symbolic domains.

Once again, Kleene confuses these two meanings in his third piece of
evidence for Church's Thesis, headed (C): Turing's concept of a
computing machine \cite[page 320]{Kleene52} where Kleene writes
``Turing's computable functions (1936-7) are those which can be
computed by a machine of a kind which is designed, according to his
analysis, to reproduce all the sorts of operations which a human
computer could perform, working according to preassigned
instructions.'' As we have seen above ``Turing's computable
functions'' are, by definition, numerical, while a human computer
faces no such restriction.

In more compressed form, this confusion re-appears in Kleene's
statement of Turing's Thesis. It is given in
Section~70.\ Turing's thesis, as a sub-ordinate clause of the opening
statement which, when elevated to an independent thesis, becomes:

\begin{description}
\item[(TT)]
  Every function which would naturally be regarded as computable is  
  computable under Turing's definition, i.e. by one of   his machines.
\end{description}

\noindent
Now the phrase ``every function which would naturally be regarded as
computable'' surely includes all computations in formal systems such
as $\lambda$-calculus or combinatory logic. For example, it would be a
distortion to assume that ``naturally'' here refers to the natural
numbers. On the other hand, ``Turing's definition'' is certainly numerical. 
As with Church's Thesis, the solution is to create a numerical thesis
(NTT) and a symbolic one (STT) as follows:

\begin{description}
\item[(NTT)]
  Every numerical function which would naturally be regarded as 
   computable is computable under Turing's definition.
\end{description}

\begin{description}
\item[(STT)] Every function which would naturally be regarded as
  computable can be simulated by a function which is computable under
  Turing's definition.
\end{description}

\noindent
From the viewpoint of Post \cite{post}, both versions of the thesis
will qualify as scientific laws. Now we can express some classical
results in the new terminology.

\begin{theorem}%[NCTT]
  Turing's Numerical Thesis is logically equivalent to Church's Numerical Thesis.
\end{theorem}
\begin{proof}
  Apply Kleene's 30th theorem, i.e.\ Theorem XXX \cite[page 376]{Kleene52}.
\end{proof}
%\begin{quote}

\begin{theorem}
\label{thm:recursive-Turing}
  The recursive model of computability is weakly equivalent to any
  Turing model of computability. 
\end{theorem}

\begin{proof}
  The traditional simulations yield encodings that are computable. 
\end{proof}

\begin{corollary}%[SCTT]
  Turing's Symbolic Thesis is logically equivalent to Church's Symbolic Thesis.
\end{corollary}
\begin{proof}
  Any simulation into a Turing model yields a simulation into the
  recursive model by composing with the simulation given by weak
  equivalence. The converse is similar.
\end{proof}

\begin{corollary}
\label{cor:lambda-inequivalent-Turing}
No $\lambda$-model of higher-order programming is weakly equivalent to the
Turing model.
\end{corollary}

\begin{proof}
  Since weak equivalence is transitive, the result follows from Corollaries~\ref{cor:lambda-inequivalent} and~\ref{thm:recursive-Turing}. 
\end{proof}

Now any reasonable notion of equivalence must imply weak equivalence.
So it follows that the Turing model of computability is strictly more
expressive than any $\lambda$-model of computability suitable for modeling
higher-order programming.

\section{The Church-Turing Thesis}
\label{sec:Church-Turing}

The mathematical confusions are now defused, with separate numerical and
symbolic versions of Church's Thesis and of Turing's Thesis, and a
clear account of equivalence of models. In turn, this must require two
versions of the Church-Turing Thesis. Putting this defusion to one
side, there remains some confusion about what, exactly, the
Church-Turing Thesis is since, although introduced in Kleene's book
\cite[page 382]{Kleene52}, it is nowhere defined. We have evidence
for four accounts.

Among the statements in Kleene's book, the closest candidate is the
opening of Section~70:
\begin{quote}
  Turing's thesis that every function which would naturally be
  regarded as computable is computable under his definition, i.e. by
  one of his machines, is equivalent to Church's thesis by Theorem
  XXX.
\end{quote}
On first reading, it is rather difficult to determine the nature of
this declaration, as it contains the statement of Turing's thesis (TT) 
plus the statement of a theorem, 
\begin{quote}
Turing's Thesis is equivalent to Church's Thesis.
\end{quote}
with its proof ``by Theorem~XXX''.  If this is the Church-Turing
Thesis, then it is a theorem asserting logical equivalence of two
theses. The other candidate statement is the numerical version of $(2)$
from Figure~\ref{fig:faulty} which is also a theorem, but this time
asserting mathematical equivalence of two models.

According to Solomon Feferman \cite{Feferman06}, the Church-Turing
Thesis was born in Alonzo Church's 1937 review of Alan Turing's paper
on computability \cite{Turing36} which declared
\begin{quote}
  As a matter of fact, there is involved here the equivalence of three
  different notions: computability by a Turing machine, general
  recursiveness in the sense of Herbrand-G\"odel-Kleene, and $\lambda$-
  definability in the sense of Kleene and the present reviewer.
\end{quote}
If by ``notion'' is meant a model of computability, then Church's
statement is about equivalence of models. Feferman goes on to say
\begin{quote}
Thus was born what is now called the Church-Turing Thesis, according
to which the effectively computable functions are exactly those
computable by a Turing machine.  The (Church-)Turing Thesis \ldots 
\end{quote}
Now Feferman identifies the Church-Turing Thesis with the
(Church)-Turing Thesis with Kleene's account of Turing's Thesis. It
seems that Feferman considers this to be a single thesis with two (or
three) names.

Finally, the literature of the last fifty years has thrown up many
versions of the theses, e.g.\ \cite{gandy, Soare96,BarendregtM13}. The
best way to make sense of this variety is to view the Church-Turing
Thesis as the class of all statements that are logically equivalent to
Church's Thesis or to Turing's Thesis.  In this manner, all of the
theses and proofs of logical equivalence are gathered under a single
heading.  This broad interpretation may explain why Kleene did not
give a statement of it.  In any event, this broad interpretation seems
most appropriate when considering the impact of the Church-Turing
Thesis on programming language design.

\section{Programming language design} 
\label{sec:design}

Here are four examples, from the last fifty years, of how confusion in the
Church-Turing Thesis has limited the design space for programming
languages.

Peter Landin's seminal paper of 1966 {\em The Next 700 Programming Languages}
\cite{Landin66} proposes a powerful model of programming language
development in which $\lambda$-calculus is the universal intermediate
language.  That is: create a source language with various additional
features such as types, or let-declarations; transform this into
$\lambda$-calculus; then implement an evaluation strategy for $\lambda$-calculus
(e.g.\ lazy or eager) as a Turing machine.  His main comment about the
suitability of  $\lambda$-calculus for this role is:
\begin{quote}
A possible first step in the research program is 1700 doctoral
theses called "A Correspondence between $x$ and Church's $\lambda$-notation.''
\end{quote}
which is footnoted ``A not inappropriate title [for this paper] would
have been ``Church without lambda.''''  So, Landin sees a central role
for $\lambda$-calculus, with a research program that would occupy a
generation of computer scientists.  The simplest
interpretation of ``Church without lambda'' is ``Church's Thesis
without lambda''.  This research program influenced many languages
designs, including Algol68 \cite{Rey81}, Scheme
\cite{sussman-steele75}, and ML
\cite{Tofte:89:TheDefinitionOfStandardML}.

Matthias Felleisen, in his paper of 1990 on the expressive power of
programming languages \cite{Felleisen90onthe}, comments:
\begin{quote}
Comparing the set of computable functions that a language can
represent is useless because the languages in question are usually
universal; other measures do not exist.
\end{quote}
After this appeal to the Church-Turing Thesis, the paper goes on to
consider various $\lambda$-calculi, in the belief that nothing has been
left out. Although Felleisen makes some useful distinctions, his paper
excludes the possibility of going beyond $\lambda$-calculus, which limits
its scope.

Henk Barendregt et al \cite{BarendregtMP13} present a current version
of the Church-Turing Thesis:
\begin{quote}
{\bf Church-Turing Thesis}
The notion of intuitive computability is exactly captured by
$\lambda$-definability or by Turing computability.
\end{quote}
It overstates the significance of $\lambda$-calculus, since ``exactly
captured'' suggests an equivalence of models, that goes beyond the
existence of mutual simulations.  It is a paraphrase of statement
$(2)$ from the faulty argument.

Robert Harper's book {\em Practical Foundations for Programming
Languages} \cite{Harper12} contains a version of Church's Thesis
(given as Church's Law) which is careful to limit its scope to natural
numbers.  It also emphasises that equality is not $\lambda$-definable
(given as Scott's Theorem). However, while the title proclaims the
subject matter to be the foundation for programming languages in
general, it is solely focused on the $\lambda$-calculus, with no allowance
made for other possibilities.  If there
remains any doubt about the author's views about foundations, consider
the following slogan, attributed to him by Dana Scott in the year of
the book's publication.  In a talk during the Turing Centenary
celebrations of 2012, he asserts \cite{Scott12}:
\begin{center}
$\lambda$ conquers all!
\end{center}
There is no explicit justification given for this focus, which we can
only assume is based upon Landin's research program, and 
the Church-Turing Thesis.

Here are some examples of calculi and languages that don't easily fit
into Landin's program, since they may exceed the expressive power of
$\lambda$-calculus.  Candidates include: first-order languages such as SQL
\cite{sql}; languages without an underlying calculus, such as Lisp
\cite{McCarthy78a} with its operators @car@ and @cdr@; and the
intensional programming language Rum \cite{Talcott:phd85}.  Richer
examples include the {\em self-calculus} of Abadi and Cardelli for
object-orientation \cite{AC96}, the {\em pattern calculus} for
structure polymorphism \cite{Jay04b}, the {\em pure pattern calculus}
for generic queries of data structures \cite{JK09, pcb}. Again, the
{\bf bondi}\ programming language \cite{bondi}  uses pure pattern
calculus to support both generic forms of the usual database queries,
and a pattern-matching account of object-orientation, including method
specialisation, sub-typing, etc. Most recently, {\em $SF$-calculus} is
a combinatory calculus that extends generic queries from data
structures to functions of all kinds \cite{JGW11}. The approach
supports definable equality of closed normal forms, typed
self-interpreters \cite{JayPalsberg11} (see also
\cite{DBLP:conf/csl/Polonsky11}), and Guy Steele's approach
\cite{steele1999growing} to growing a language \cite{jay2013growing}.
It can also be extended to a concurrent setting
\cite{DBLP:journals/corr/Given-WilsonGJ14,GivenWilson14}.
The richer calculi above have not been shown equivalent to $\lambda$-calculus.
Rather, all evidence points the other way, since the factorisation
operator $F$ of $SF$-calculus cannot be defined in $SK$-calculus
\cite{JGW11}.

Summarising, while the ability to simulate $\lambda$-calculus as a Turing
machine has been enormously fruitful, the larger claims of the
Church-Turing Thesis have been suggesting unnecessary limits on
programming language design for almost fifty years.

\section{Intensional computation} 
\label{sec:intensional}

Having exposed a gap between the expressive power of $\lambda$-calculus and
of Turing machines, it is natural to consider how to bridge it. 

It is not a simple matter to overcome the limitations of $\lambda$-calculus
by, say, adding an operator for equality.  The essential difficulty is
that while $\lambda$-terms describe algorithms, i.e.\ capture {\em
  intensions}, this intensional information cannot be recovered from
within $\lambda$-calculus in any uniform manner.  Rather $\lambda$-terms can
only extract {\em extensional} information about input-output
behaviour \cite[page 2]{Church41}.  To redress this, various efforts
have been made, in the context of partial evaluation and
decompilation, to extend $\lambda$-calculus with G\"odelisation.  This has
been done for simply typed $\lambda$-calculus
\cite{Berger-Schwichtenberg91}, a combinatory calculus
\cite{goldberg2000godelization}, and untyped $\lambda$-calculus augmented
with some labels \cite{goedelization:mogensen:PEPM99}.  However, some
of the attractive features of pure $\lambda$-calculus, such as being
typable, confluent, and a rewriting system have been compromised.

Alternatively, the development of intensional computation can begin
afresh.  Intensionality has been the subject of much research by
% materna09
philosophers \cite{Montague60,feferman60,cellucci81,katz01}, logicians
% Dybjer1996
\cite{frege1892,tichy69,Brouwer81}, type theorists \cite{lof,muskens2007,bove2009brief} and computer scientists
% gilmore01,fox02
\cite{hobbs78,ArtemovB07,carette12}, so before proceeding, let us
determine what it will mean for us. In the concrete setting of
$\lambda$-calculus, when should two $\lambda$-terms be considered intensionally
equal? Should this be limited to closed normal forms or are arbitrary
terms to be included? In part, the answer depends upon whether your
semantics is denotational or operational.

{\em Denotational semantics} constructs the meaning of a program from that
of its fragments, whose contexts may supply values to free variables,
or determine whether or not the evaluation of the fragment terminates.
Examples may be found in {\em domain theory} \cite{Kreisel69,
  Scott76}, {\em abstract and algebraic data types} \cite{ADJ77,
  Mitchell88}, the {\em effective topos} \cite{Hyland82}, and {\em
  partial combinatory algebras} \cite{pca, CockettH08,Longley14}.
%
%However, no such model of computability can include an equality
%function, as this would yield a solution to the Halting Problem
%\cite[Section 17.3]{Harper12}. 
%
This suggests that all terms are included but equality of arbitrary
lambda terms is not computable \cite[page 519]{Bare84a}.
% or Harper: [Section 17.3].  

By contrast, other semantics do not account for terms without normal
form or for open terms, and so avoid the need to assign them values.
For example, {\em axiomatic recursion theory} \cite{Rogers67, Jones97}
uses Kleene equality \cite[page 327]{Kleene52}.  Again, {\em operational semantics} in the style
of Gordon Plotkin's {\em structured operational semantics}
\cite{Plotkin04} can limit its values to be closed terms that are, in
some sense, normal, e.g.\ are irreducible, or in head-normal form
\cite{Bare84a}, etc. Thereby, various problems caused by non-termination,
such as the difficulty of defining the parallel-or function
\cite{Abramsky00}, do not arise. In particular, it is easy to
represent equality.

Thus, the challenge is to extend standard calculi with the ability to
query the internal structure of closed normal forms, in a process akin
to G\"odelisation, while retaining many of the attractive features of
$\lambda$-calculus, such as being a rewriting system, especially one that
is confluent or typable.

The $SF$-calculus achieves this by replacing the operator $K$ of
$SK$-calculus with a {\em factorisation operator} $F$ that is able to
test the internal structure of terms that are, in some sense, head
normal, by factoring them.  The operator $F$ takes three arguments. If
$O$ is an operator and $M$ and $N$ are combinators then $FOMN$ reduces
to $M$, so that $FF$ represents the traditional $K$. If $PQ$ is a {\em
  compound} then $F(PQ)MN$ reduces to $NPQ$, so that $N$ can
manipulate the components separately. It cannot be stressed too much
that not every application is a compound.  Rather, compounds are
applications that can never be reduced at their head.  That is, they
are given by all partially applied operators such as $SM$ and $SMN$
and $F$ and $FMP$ for any terms $M, N$ and $P$.  Non-examples include
fully applied operators such as $SMNP$ and $FMNP$, and terms headed by
a variable, such as $xM$. The latter is excluded since substitution
may create a fully applied operator. Using factorisation, closed
normal forms can be completely analysed into their constituent
operators, whose equality can be tested by extensional means.  Details of the calculus
can be found in the original paper \cite{JGW11}.

\begin{theorem}
  The normal model of computability for $SF$-calculus is equivalent
  to the recursive model. 
\end{theorem}

\begin{proof}
  That G\"odelisation and Church encoding are both simulations follows
  from the work of Church \cite{Church36} and Kleene \cite{Kleene36},
  so that it is enough to show that both re-codings are computable. It
  is easy to see that the recoding of numbers to numbers is recursive.
  In the other direction, the recoding of $SF$-combinators can be
  described by a pattern-matching function that acts on the
  combinators in normal form.  Such pattern-matching functions are
  represented by $SF$-combinators because $SF$-calculus is {\em
    structure complete} \cite{JGW11}. Note that this proof does not
  apply for $SK$-calculus as this is merely {\em combinatorially
    complete} \cite{TRS}. Also, since the recodings are invertible,
  this is an equivalence of partial combinatory algebras, in the sense
  of Longley \cite{Longley14}.
\end{proof}

To the extent that $\lambda$-calculus is equivalent to $SK$-calculus, this
makes $SF$-calculus a superior foundation for higher-order programming
languages.

\section{Conclusions}
\label{sec:conclusions}

The Church-Turing Thesis has been a confusion since it was first
named, but not defined, by Kleene in 1952.  The numerical results of
Church and Turing support numerical versions of their eponymous theses, in which
sets of numerical functions are co-extensive. Further, there are
separate theses for symbolic computation, involving simulations of one
model of computability in another. Kleene confused these two
settings, with a little encouragement from Church.  

Once the role of simulations is made explicit, it is easier to see
that mutual simulation yields an equivalence of models only if both
re-codings are computable, each in its respective model. This
requirement exposes the limitations of $\lambda$-calculus, since
G\"odelisation is not $\lambda$-definable, even for closed $\lambda$-terms in
normal form.

These limitations are, in some sense, well known within the
$\lambda$-calculus community, in that $\lambda$-calculus cannot define equality,
even of closed normal forms. Indeed, those working with categorical
models of computability, or analysing programs defined as $\lambda$-terms,
are acutely aware of these limitations. However, the community as a
whole is not keen to advertise any of this, proclaiming instead that
$\lambda$ conquers all. Students who ask the wrong questions may be told
``Beware the Turing tarpit!''  \cite{Perlis82} or ``Don't look under
the lambda!''  which closes off discussion without clarifying anything.

The limitations of $\lambda$-calculus are essential to its nature, since
$\lambda$-terms cannot directly query the internal structure of their
arguments; the expressive power of $\lambda$-calculus is extensional. This
does not matter for numerical computations since the internal
structure of a natural number is determined by the zero-test and the
predecessor function, both of which are recursive.  However, this
approach cannot be generalised to query internal structure in richer
settings.

Rather, intensional computation requires a fresh outlook. The simplest
illustration of this is the $SF$-calculus whose factorisation operator
$F$ is able to uniformly decompose normal forms to their constituent
operators.  Since $SF$-calculus also has all of the expressive power
of $SK$-calculus, its normal model of computability is equivalent to
the Turing model or the recursive function model.

The implications of this for programming language design are profound.
The \bondi\ programming language has already shown how the usual
database queries can be made polymorphic, and that object-orientation
can be defined in terms of pattern-matching.  Now the factorisation
operator paves the way for program analysis to be conducted in the
source language, so that growing a language can become easier than
ever.

In short, confusion in the Church-Turing Thesis has obscured the
fundamental limitations of $\lambda$-calculus as a foundation for
programming languages.  It is time to wind up Landin's research
program, and pursue the development of intensional calculi and
programming languages.

\medskip

\noindent {\bf Acknowledgments} We thank Jacques Carette, Robin Cockett, John Crossley,
Nachum Dershowitz, Thomas Given-Wilson, Neil Jones, Jens Palsberg, Reuben Rowe,
Peter Selinger, and Eric Torreborre for comments on drafts of this paper.

\bibliographystyle{plain} % {ACM-Reference-Format-Journals}
\bibliography{Jay,foundations}

\newpage 
\tableofcontents

\end{document}